\documentclass{article}
\usepackage{hiph-art}
\volnumber{19} \issuenumber{1} \edyear{2004}                             
\frompage{000} \topage{000}                                              
\recrevdate{1 January 2004}                                              
\def\rightmark{NeXSPheRIO results}
\def\leftmark{R.Andrade et al.}
 
\title{NeXSPheRIO results on elliptic flow at RHIC 
and connection with thermalization }
\authors{
{R.Andrade$^1$, \underline{F.Grassi}$^1$, 
Y.Hama$^1$, T.Kodama$^2$,
O.Socolowski Jr.$^3$,
and B.Tavares$^2$
\index{Andrade R.} 
\index{Grassi F.} 
\index{Hama Y.}
\index{Kodama T.}
\index{Socolowski Jr. O.}
\index{Tavares B.}
}\\[2.812mm]
{\normalsize
\hspace*{-8pt}$^1$ Instituto de F\'{\i}sica, USP,\\ 
C. P. 66318, 05315-970 S\~{a}o Paulo-SP, Brazil\\[0.2ex] 
\hspace*{-8pt}$^2$ Instituto de F\'{\i}sica, UFRJ, \\ 
C. P. 68528, 21945-970 Rio de Janeiro-RJ , Brazil\\[0.2ex]
\hspace*{-8pt}$^3$ CTA/ITA,\\
Pra\c{c}a Marechal Eduardo Gomes 50, 
CEP 12228-900 S\~ao Jos\'e dos Campos-SP, Brazil
}}
 
\abstract{
Elliptic flow at RHIC is computed event-by-event
with NeXSPheRIO. Reasonable agreement with experimental results on
$v_2(\eta)$ is obtained. Various effects are studied as well:
reconstruction of impact parameter direction, freeze out temperature,
equation of state (with or without crossover), emission mecanism.}


 
\makeindex
\begin{document}
 
\maketitle

\section{Motivation}\label{motiv}

Hydrodynamics seems a correct tool to describe RHIC collisions however 
$v_2(\eta)$ is not well reproduced  as shown by
Hirano et al. \cite{hi}. These anthors suggested
that  this might be due to lack of thermalization.
Heinz and Kolb\cite{hk} presented a model with partial thermalization
and obtained a reasonable agreement with data.
The question addressed in this work is whether
lack of thermalization is the only explaination for this disagreement between 
data and theory for $v_2(\eta)$.

\section{Brief description of NeXSPheRIO}

The tool we use is the hydrodynamical code called
NeXSPheRIO. It is a junction of two codes.

The SPheRIO code   is used to compute the hydrodynamical evolution. It is based on
Smoothed Particle Hydrodynamics, a method
originally developped in astrophysics
and adapted to relativistic heavy ion
collisions \cite{spherio}.
Its main advantage is that
 any geometry in the initial conditions can be incorporated.

The  NeXus code  is used to compute the initial conditions
$T_{\mu \nu}$, $j^{\mu}$ and $u^{\mu}$ on a proper time hypersurface \cite{IC}.
An example of  initial condition for one event is shown in figure 1.

\begin{figure}[htb]
                 \insertplot{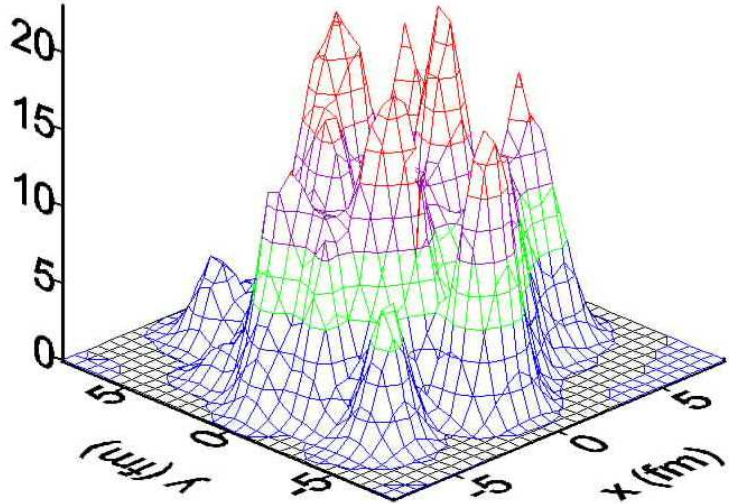}
\caption[]{Example of initial energy density in the $\eta=0$ plane.}
\label{fig1}
\end{figure}

NeXSPheRIO is run many times, corresponding to many different events or initial conditions. In the end,
 an average over  final results is performed.
This mimicks experimental conditions.
This is different from the canonical approach in hydrodynamics where
initial conditions are adjusted to reproduce some selected data and are very 
smooth.

This code has been used to study a range of problems concerning relativistic nuclear collisions: effect of fluctuating initial conditions on particle 
distributions \cite{FIC}, energy dependence of the kaon effective temperature 
\cite{kaon}, interferometry at RHIC \cite{HBT}, transverse mass ditributions 
at SPS for strange and non-strange particles \cite{strange}.

\section{Results}

\subsection{Theoretical vs. experimental computation}

Theoretically, the impact parameter angle $\phi_b$ is known
and varies in the range of the centrality window chosen. The elliptic flow can be computed easily through 
\begin{equation}
<v_2^b(\eta)>=<\frac{\int d^2N/d\phi d\eta \cos[2(\phi-\phi_b)]\, d\phi}
{\int d^2N/d\phi d\eta \, d\phi}>
\end{equation}
The average is performed over all events in the centrality bin.
This is shown by the lowest solid curve in figure 2.

Experimentally, the impact parameter angle $\psi_2$ is reconstructed and
a correction is applied to the elliptic flow computed with respect to this angle, 
to correct for the reaction plane resolution.
For example in a Phobos-like way \cite{phobos}
\begin{equation}
<v_2^{b,rec}(\eta)>=<\frac{v_2^{obs}(\eta)}
         {\sqrt{<\cos[2(\psi_2^{<0}-\psi_2^{>0})]>}}>
\end{equation}
where
\begin{equation}
v_2^{obs}(\eta)=\frac{\sum_i d^2N/d\phi_i d\eta \cos[2(\phi_i-\psi_2)]}
          {\sum_i d^2N/d\phi_i d\eta}
\end{equation}
and
\begin{equation}
\psi_2=\frac{1}{2} \tan^{-1} \frac{\sum_i \sin 2 \phi_i}{\sum_i \cos 2 \phi_i}
\end{equation}

In the hit-based method, 
$\psi_2^{<0}$ and $\psi_2^{>0}$ are determined for subevents 
$\eta < 0$ and $>0$ respectively and
if $v_2$ is computed for a positive (negative) $\eta$,
the sum in $\psi_2$, equation 3,  is over particles with $\eta < 0$ ($\eta > 0$).

In the track-based method, 
$\psi_2^{<0}$ and $\psi_2^{>0}$ are determined for subevents 
$2.05<\mid \eta \mid < 3.2$ and 
$v_2$ is obtained for particles around $0<\eta < 1.8$ and reflected
(there is also an additional $\sqrt{2}$ in the reaction plane correction
in equation 2).

In figure 2, we also show the results for $v_2^{obs}(\eta)$ 
for both the hit-based  (dashed line) and
track-based (dotted line) methods.
We see that both curves can lie {\em above} the theoretical $<v_2^b(\eta)>$
 (solid)  curve.
So dividing them by a cosine to get $<v_2^{b,rec}(\eta)>$
will make the disagreement worse:  $<v_2^{b}(\eta)>$ and 
$<v_2^{b,rec}(\eta)>$ are different.

\begin{figure}[htb]
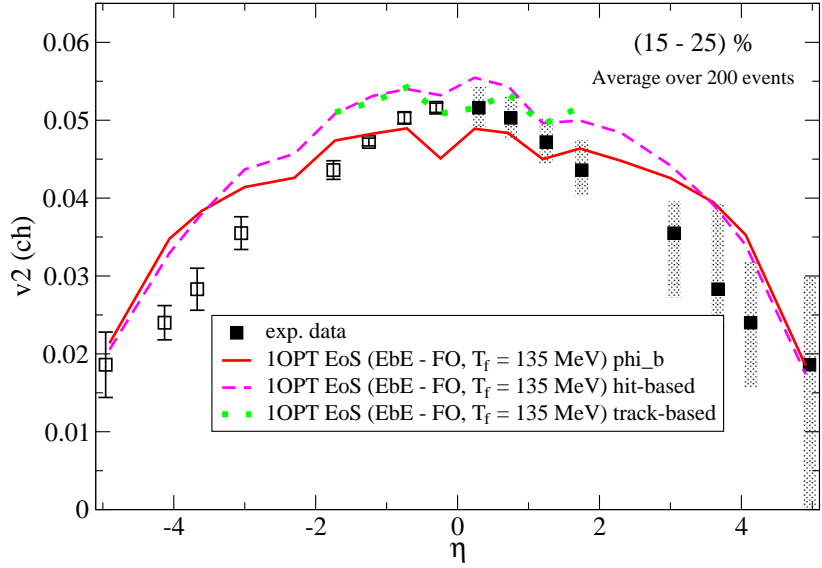

                 \insertplot{POSpresv2_ch_eta_j3_2.eps}
\caption[]{Comparison of various ways of computing $v_2$: solid line is 
using 
the known impact parameter angle $\phi_b$, dashed  and dotted lines is using the 
reconstructed impact parameter angle $\psi_2$.  1OPT stands for equation of state with first order transition, 
EbE, event-by-event calculation, FO, freeze out mechanism for particle 
emission.
Data are from Phobos 
\cite{phobos}.
For more details see text.}
\label{fig2}
\end{figure}

Since the standard way to 
include the correction for the reaction plane resolution (equation 2)
seems inapplicable,
 we need to understand why.
When we look at the distribution $d^2N/d\phi d\eta$ obtained with NeXSPheRIO,
it is not symmetric with respect to the reaction plane.
This happens because the number of produced particles is finite. Therefore we 
must write
\begin{eqnarray}
\hspace*{-0.5cm}
\frac{d^2N}{d\phi d\eta}& =& v^b_0(\eta) [1+ \sum 2 v^b_n(\eta) \cos(n(\phi-\phi_b))+ 
\sum 2 v'^{b}_n(\eta) \sin(n(\phi-\phi_b)) ]\\
& = & v^{obs}_0(\eta) [1+ \sum 2 v^{obs}_n(\eta) \cos(n(\phi-\psi_2))+ 
\sum 2 v'^{obs}_n(\eta) \sin(n(\phi-\psi_2)) ]
\end{eqnarray}
It follows that
\begin{equation}
v_2^{obs}(\eta)=v_2^b(\eta) \cos[2(\psi_2-\phi_b)] 
+ v'^b_2(\eta) \sin[2(\psi_2-\phi_b)]
\end{equation}
We see that due to the term in sine, we can indeed have
$<v_2^{obs}(\eta)>$ larger than $<v_2^b(\eta)>$,
 as in  figure 2. (The sine term does not vanish upon averaging on events because if a choice such as equation 4 is done for $\psi_2$,
$v'^b_2(\eta)$ and $\sin(2(\psi_2-\phi_b)$ have same sign. Rigorously,
this sign condition
is true if $\psi_2$ is computed for the same $\eta$ as $v'^b_2(\eta)$.
Due to the actual way of extracting $\psi_2$ experimentally, 
we expect this condition is  satisfied for particles with small  or moderate pseudorapidity.)
In the standard approach, it is supposed that 
$d^2N/d\phi d\eta$  is symmetric with respect to the reaction plane
 and there are no sine terms in the Fourier
decomposition
of $d^2N/d\phi d\eta$ (equation 5); 
as a consequence, $v_2^{obs}(\eta)\leq v_2^b(\eta)$.

Since the experimental results for elliptic flow are obtained assuming that 
$d^2N/d\phi d\eta$ is symmetric around the reaction plane, 
we cannot expect perfect agreement of our 
$<v_2^b(\eta)>$ with them. 
In the following we use the theoretical method, i.e. $<v_2^b(\eta)>$,
 to make further comparisons.

\subsection{Study of various effects which can influence the shape of 
$v_2(\eta)$}

In all comparisons, the same set of initial conditions is used, scaled to reproduce 
$dN/d\eta$ for $T_{f.out}=135$ MeV. 

First we study the effect of the freeze out temperature on
the pseudo-rapidity
and transverse momentum distributions as well as $v_2(\eta)$ 
(this last quantity is shown in figure 3).
We found that
 $v_2(\eta)$ and $d^2N/p_t\,dp_t$
favor $T_{f.out}=135$ MeV, so this temperature is used thereafter.
\begin{figure}[htb]
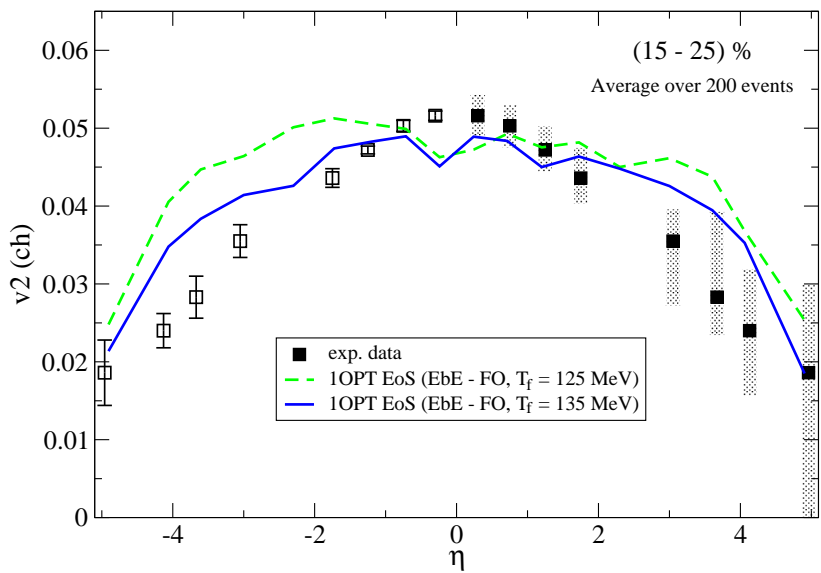

\vspace*{1.cm}
\insertplot{POSpresd2n_ch+-_dy_pi_dpt_j3_3.eps}
\caption[]{Comparison of $v_2(\eta)$
for two freeze out temperatures. Abbreviations: see figure 2.}
\label{fig3}
\end{figure}

 We now compare results obtained for a quark matter equation of state with
first order transition to hadronic matter and with a crossover
(for details see 
\cite{YHqm2005}).
We have checked that the $\eta$ and $p_t$ distributions are not much affected.
We expect larger $v_2$ for cross over because there is always acceleration
and this is indeed what is seen in figure 4.
\begin{figure}[!b]
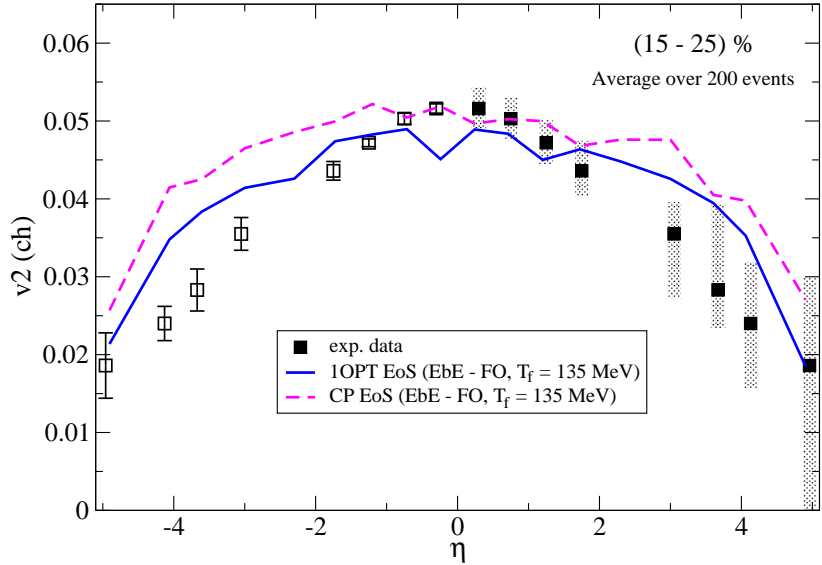

\vspace*{1.5cm}
                 \insertplot{POSpresv2_ch_eta_j3_5.eps}
\caption[]{Comparison of  $v_2(\eta)$
for first order transition (1OPT) and critical point (CP) equations of state.}
\label{fig5}
\end{figure}

 We then compare results obtained for freeze out and continuous emission \cite{CEM}.
Again,
we have checked that the $\eta$ and $p_t$ distributions are not much affected.
We expect  earlier emission, with less flow, at large $|\eta|$ regions,
 therefore, narrower $v_2(\eta)$ 
and this is indeed what is seen in figure 5.

Finally, we note that
compared to Hirano's pioneering work with smooth initial conditions, 
the fact that we used event-by-event initial conditions seems crucial: we 
immediately avoid the two bump structure. To check this, it is interesting to
 study what we would  get with smooth initial conditions. We obtained such conditions by averaging the initial conditions of 30 Nexus events.
Again,
we have checked that the $\eta$ and $p_t$ distributions are not much affected but preleminary results shown in figure 6 indicate that
now $v_2$ is very different, having
 a  bumpy structure. The case of smooth initial conditions has a well defined
asymmetry and the elliptic flow reflects this. 
The ellipict flow of the 
event-by-event case is an average over results obtained for randomly varying 
initial conditions, each with a different asymmetry. As a consequence,
the average $v_2$ has a smoother behavior but  
large fluctutations \cite{YHqm2005} and is smaller 
(around the initial energy density sharp
peaks seen in figure 1, in each event,
expansion is more symmetric. No such sharp peak exists for the average initial conditions).

\section{Summary}

$v_2(\eta)$ was computed with NeXSPheRIO at RHIC energy.
 Event-by-event initial conditions seem important to get the right shape
of $v_2(\eta)$ at RHIC. Other features seem  less important:
freeze out temperature, 
equation of state (with or without crossover), emission mechanism.
Finally, we have shown that the reconstruction of the impact parameter direction $\psi_2$, as given by eq. (4), gives $v_2^{obs}(\eta)>v_2^b(\eta)$,
when taking into account the fact that the azimuthal particle distribution is 
not 
symmetric with respect to the reaction plane.

Lack of thermalization is not necessary to reproduce
$v_2(\eta)$. The fact that there is thermalization outside mid-pseudorapidity
is reasonable given that the (averaged) initial energy density is high
 there
(figure not shown).
A somewhat similar conclusion was  obtained by Hirano at this 
conference,
using color glass condensate initial conditions for a hydrodynamical code and emission 
through a cascade code \cite{hiqm05}.

\vspace*{0.5cm}
We acknowledge financial support by FAPESP (2004/10619-9, 2004/13309-0,
2004/15560-2, CAPES/PROBRAL, CNPq, FAPERJ and PRONEX.


\newpage

\begin{figure}[htb]
\vspace*{0.3cm}
                 \insertplot{POSpresv2_ch_eta_j3_10.eps}
\vspace*{-1.1cm}
\caption[]{Comparison of  $v_2(\eta)$
for freeze out (FO) and continuous emission (CE).}
\label{fig6}
\end{figure}

\begin{figure}[htb]
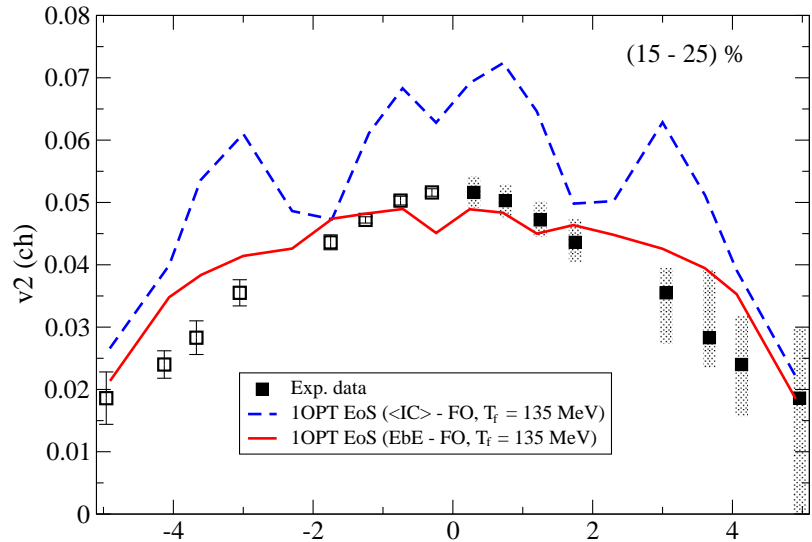

\vspace*{0.1cm}
                 \insertplot{POSpresv2_ch_eta_j3_7.eps}
\vspace*{-1.1cm}
\caption[]{Comparison of  $v_2(\eta)$
computed event-by-event (EbE) and with smooth initial conditions ($<IC>$).}
\label{fig7}
\end{figure}

\end{document}